\documentclass[usenatbib]{mn2e}

\usepackage[]{graphicx}

\begin{document}

\title{PSR J1022+1001: Profile Stability \& Precision Timing} 
\author[A. W. Hotan, M. Bailes and S. M. Ord.]
       {A.~W.~Hotan,$^{1,2}$
        M.~Bailes,$^1$ 
	S.~M.~Ord$^1$
\\
\\
$^1$Swinburne University of Technology,\\
    Centre for Astrophysics and Supercomputing \\
    Mail \#31
    P. O. Box 218
    VIC 3122
    Australia
\\
\\
$^2$Australia Telescope National Facility, \\
    PO Box 76
    Epping NSW 1710
    Australia
}

\maketitle

\begin{abstract}

We present an investigation of the morphology and arrival times of
integrated radio pulses from the binary millisecond pulsar PSR
J1022+1001. This pulsar is renowned for its poor timing properties,
which have been postulated to originate from variability in its
average pulse profile. Although a sub-class of long-period pulsars are
known to exhibit mode changes that give rise to very large deviations
in their integrated profiles, this was the first millisecond pulsar
thought to have an unstable mean profile. As part of a precision
timing program at the Parkes radio telescope we observed this pulsar
between January 2003 and March 2004 using a coherent de-dispersion
system (CPSR2). A study of morphological variability during our
brightest observations suggests that the pulse profile varies by at
most a few percent, similar to the uncertainty in our calibration.
Unlike previous authors, we find that this pulsar times extremely
well. In five minute integrations of 64 MHz bands we obtain a weighted
RMS residual of just 2.27 $\mu$s. The reduced $\chi^{2}$ of our best
fit is 1.43, which suggests that this pulsar can be timed to high
accuracy with standard cross-correlation techniques. Combining
relativistic constraints with the pulsar mass function and
consideration of the Chandrasekhar mass limit on the white dwarf
companion, we can constrain the inclination angle of the system to lie
within the range $37^o < i < 56^o$. For reasonable pulsar masses, this
suggests that the white dwarf is at least 0.9 M$_\odot$. We also find
evidence for secular evolution of the projected semi-major axis.

\end{abstract}

\section{Introduction}
\label{sec:intro}

Pulsar timing is a highly versatile experimental technique that is
used to provide estimates of everything from the age and magnetic
field strengths of pulsars, allowing population studies
\citep{lmt85,lbdh93}, to micro-arc-second positions and tests of
relativistic gravity \citep{vbb+01}. Central to the technique is the
maxim that pulsars have average pulse profiles that are inherently
stable \citep{hmt75}, which can be cross-correlated with a template to
provide accurate times of arrival \citep{mt77}. Once corrections are
made to account for the Earth's position with respect to the solar
system barycentre, a model can be fit to the data, transforming
pulsars from astrophysical curiosities into powerful probes of binary
evolution and gravitational theories. One spectacular example is the
recently-discovered double pulsar system \citep{lbk+04}.

The discovery of the millisecond pulsars PSR B1937+21 and B1855+09
\citep{bkh+82, srs+86} and their subsequent timing \citep{ktr94}
revealed that millisecond pulsars were very rotationally stable,
rivaling the best atomic clocks. Their lower magnetic fields appeared
to give rise to a fractional stability that far exceeded that of the
normal pulsar population \citep{antt94}. This stability, combined with
their small rotation periods, gives the millisecond pulsars
sub-microsecond timing potential provided sufficient signal-to-noise
ratio (S/N) pulse profiles can be recorded. This provides interesting
limits on a wide range of astrophysical phenomena \citep{s04}.

An array of accurately timed MSPs can even be used to constrain
cosmological models and search for long-period (nHz) gravitational
waves \citep{fb90}. Such an array requires frequent observations of a
selection of pulsars that are known to time to a precision of less
than 1 $\mu$s, preferably in two or more radio frequency
bands. Implementation of such an array would provide an opportunity to
directly detect gravitational radiation in a regime that complements
the sensitivity range of ground-based detectors, to test the accuracy
of solar system ephemerides and to construct an entirely
extra-terrestrial time scale. The question of whether a suitable
sample of millisecond pulsars positioned throughout the entire
celestial sphere can be selected from the pulsar catalogue
\citep{hmt+04} is under active assessment, through a collaboration
between the Australia Telescope National Facility and Swinburne
University of Technology.

The feasibility of a timing array depends in part upon the intrinsic
rotational stability of the MSPs and their lack of pulse profile
variation. \cite{cb04} have recently discovered a ``glitch'' in the
millisecond pulsar PSR B1821--24, similar to those discovered in
younger pulsars \citep{sl96}. Perhaps more disturbingly, \cite{kxc+99}
report that the binary millisecond pulsar PSR J1022+1001 exhibits
pulse shape variations that ruin its timing. Studies conducted by
\cite{kxc+99, rk03} argue that the pulsar exhibits significant changes
in its pulse morphology on $\sim$5 minute time scales and narrow
bandwidths. They interpreted these variations as the source of the
pulsar's unusually poor timing properties. By modeling the profile
with Gaussian components, \cite{kxc+99} improved the timing and argued
that the trailing component of the pulse was more stable than the
leading component.

The precision with which astronomers can predict pulse arrival times
has been steadily improving over the past few decades with the advent
of new technology and methodology. The charged interstellar medium
disperses radio pulses and broadens the pulse profile across finite
receiver bandwidths. Astronomers crave both the wide bandwidths that
permit high S/N integrated profiles, and the resolution of sharp
features that permit strong cross-correlation with a standard template
in order to minimise timing errors.  Pulsar astronomy has always
benefited from adopting new technologies that give increased time
resolution and help defeat the deleterious effect of the interstellar
medium on the propagation of radio pulses.

Incoherent devices such as analogue filter-banks are prone to
systematic errors due to interstellar scintillation and imperfect
frequency responses. Digital spectrometers can overcome some of these
inadequacies but \cite{hr75} described a computational technique
(coherent de-dispersion) that removes pulse dispersion ``perfectly''
almost three decades ago. Unfortunately, at the time, computational
hardware was not capable of processing all but the most modest of
bandwidths.  The exponential growth of computational power has
permitted the development of new pulsar instruments that are capable
of coherently de-dispersing large bandwidths in near real time. Such
an instrument is CPSR2, a baseband recorder that permits near-real
time coherent de-dispersion of 2$\times$64 MHz bands using a cluster
of high-end workstations \citep{bai03}. These new instruments
represent the best opportunity to study small variations of integrated
pulse profiles because they deliver an immunity against systematic
errors induced by narrow-band scintillation and dispersion measure
smearing. They also provide full polarimetry and multi-bit
digitisation, radio frequency interference rejection and the
opportunity to apply accurate statistical corrections that help
eliminate systematic errors induced by discretisation\citep{ja98}.

Motivated by the arrival of CPSR2, we have commenced a timing campaign
of the best millisecond pulsars in an effort to make progress towards
the aims of the timing array. PSR J1022+1001 was included because of
its high flux density, and the paucity of better candidates at the
hour angle it is visible from the Parkes radio telescope. PSR
J1022+1001 is a recycled or millisecond pulsar with a pulse period of
approximately 16 ms. It orbits once every 7.8 days in a binary system
with a companion whose minimum mass can be derived from the pulsar
mass function, Eqn \ref{eqn:massfn}, if we assume that the orbit is
face on and the pulsar has a mass of 1.35 M$_\odot$.

\begin{equation}
f(m_{\rm p},m_{\rm WD}) = {{m_{\rm WD}^3} \sin^3
i\over{(m_{\rm p}+m_{\rm WD})^2}} = {{4 \pi^2}\over{G}}{{a^3 \sin^3
i}\over{P_{\rm b}^2}}
\label{eqn:massfn}
\end{equation} 

\noindent Here $m_{\rm p}$ and $m_{\rm WD}$ refer to the mass of the
pulsar and white dwarf respectively, $i$ is the inclination angle of
the orbit, $P_{\rm b}$ is the orbital period and $G$ is Newton's
gravitational constant.  For PSR J1022+1001, this corresponds to a
mass of at least 0.7 M$_\odot$.

Measurements of the dispersion measure along the line of site to the
pulsar, combined with the \cite{tc93} galactic electron density model,
place PSR J1022+1001 at a distance of roughly 600 pc from the sun,
making it a relatively nearby source.  Accurate timing of this pulsar
could therefore lead to a greater knowledge of its orientation,
companion masses, distance (via parallax) and proper motion.
Unfortunately, this pulsar is very near the ecliptic plane and hence
pulsar timing is only good at accurately constraining its position in
one dimension.  It is therefore a good target for very long baseline
interferometry, as this could accurately determine the position and
proper motion in ecliptic latitude.

In this paper we present an analysis of the first 15 months of CPSR2
observations of PSR J1022+1001, demonstrating that this pulsar can be
timed to high accuracy using standard techniques. In section
\ref{sec:observations} we describe the observing system and
methodology, followed by a description of our data reduction scheme,
including polarimetric calibration. Section \ref{sec:stability}
describes the results of a search within our data set for variations
of the type reported by \cite{kxc+99}. Pulse arrival times calculated
from this data set are analysed in section \ref{sec:timing}, followed
by a summary of the newly derived pulsar spin and binary system
parameters, including those that may become significant in the
future. In section \ref{sec:conclusions} we discuss the implications
of our work for this pulsar and precision timing programs in general.

\section{Observations} 
\label{sec:observations}

\subsection{Instrumentation}

The second Caltech Parkes Swinburne Recorder (CPSR2) is a baseband
recording and online coherent de-dispersion system that accepts
4$\times$64 MHz intermediate frequency bands (IFs).  Using the
in-built supercomputer, data can be processed either in real time if
the pulsar has a dispersion measure (DM) less than approximately 50 pc
cm$^{-3}$ at a wavelength of 20 cm, or recorded to disk and processed
offline. This allows a large amount of flexibility in observing
methodology. CPSR2 was commissioned in August 2002 and has been
recording data on a regular basis since about November 2002. Observing
sessions are primarily conducted using a standard configuration
consisting of two independent 64 MHz bands, each with two orthogonal
linear polarisations. The down-conversion chain is configured to make
both bands contiguous, at centre frequencies of 1341.0 and 1405.0
MHz. In addition, a small number of observations are made at the
widely separated frequencies of 3000.0 MHz and 685.0 MHz. After
down-conversion and filtering to create band-limited signals, each IF
is fed into the CPSR2 Fast Flexible Digitiser board which performs
real, 2-bit Nyquist sampling. The digitisation process generates a
total of 128 million bytes of data every second. These samples are fed
via Direct Memory Access (DMA) cards to two high-speed computers that
divide the data into discrete blocks which are distributed via gigabit
ethernet to a cluster of client machines for reduction using a
software program called PSRDISP \citep{van03}.  PSRDISP implements
software-based coherent de-dispersion \citep{hr75}, assuming prior
knowledge of the DM along the line of sight to the pulsar and the
validity of the tenuous plasma dispersion law.

\subsection{Summary of Observations}

Observations were conducted at the Parkes radio telescope over a
period of around 400 days between January 2003 and February 2004. The
central beam of the Parkes Multibeam receiver \citep{swb+96} was
originally used at the front end, providing a system temperature of
approximately 21K at 20 cm. The Multibeam receiver was removed for
maintenance in October 2003 and all 20 cm data since this date were
recorded using the refurbished, wide-band H-OH receiver whose
observable band encompasses that of the Multibeam receiver. The H-OH
system is slightly less sensitive than the Multibeam, with a system
temperature of 25K at 20 cm wavelengths. In addition, a new coaxial
dual-band 10/50 cm receiver system was installed in place of the
Multibeam. During the 2004 observing sessions, data were taken with
this system in order to expand our frequency coverage for the purposes
of dispersion measure monitoring. Preliminary system temperature
measurements of the coaxial system yield values of 30 K and 40 K for
the 10 cm and 50 cm feeds respectively.

Scheduled observing sessions were typically a few days in duration,
occurring once or twice a month. Individual tracks of PSR J1022+1001
vary from approximately 15 minutes to a few hours in duration. The
data are somewhat biased towards episodes of favourable scintillation
as this allows the most efficient use of limited telescope time. In
addition to the on-source tracks, most observations were immediately
preceded or followed by a short (2 minute) observation, taken 1 degree
south of the pulsar position, during which the receiver noise source
was driven with a square wave at a frequency of 11.122976 Hz. These
calibration tracks were used to characterise the polarimetric response
of the signal chain so that corrections to the observed Stokes
parameters could be made during data reduction. Observations of the
radio galaxy 3C218 (Hydra A) were also taken at approximately monthly
intervals (though often only at 20 cm) and used to calibrate the
absolute flux scale of the observing system.

\subsection{Data Reduction}

Data from each dual-polarisation band were reduced to a coherent
filter bank with 128 spectral channels, each 0.5 MHz wide,
corresponding to an effective sampling time of 2 $\mu$s. 1024 phase
bins were stored across the pulse profile, yielding a time resolution
for PSR J1022+1001 of about 15 $\mu$s per bin. Each individual data
block recorded by CPSR2 represents 16.7 seconds of data (1 Gigabyte of
2-bit samples). These blocks can later be combined to form integrated
profiles. For this purpose the authors use the application set
provided with the PSRCHIVE \citep{hvm04} scheme. Due to the
IERS\footnote{http://www.iers.org} practice of retrospectively
publishing corrections to the rotation rate of the Earth, the data
presented in this paper are that for which corrections to the solar
system barycentre could be accurately made at the time of
preparation. Observations of PSR J1022+1001 are ongoing.

\subsection{Processing \& Calibration}

The set of 16.7 second PSR J1022+1001 integrations were summed to a
total integration time of 5 minutes. These data were calibrated to
account for the polarimetric response of the observing system, then
all frequency channels and Stokes parameters were combined to form a
single total intensity profile corresponding to each 5 minute
integration. All of the receiver systems used during observations of
PSR J1022+1001 are equipped with orthogonal linear feed probes, so the
recorded polarisation products were corrected for relative gain and
phase. Off-source calibrator observations were used to compute the
relative gain and phase terms for each receiver system at the epoch of
the observation, using a simple case of the scheme described by
\cite{vs04}. We do not attempt to correct for more subtle errors
arising from cross-contamination between the two probes, primarily
because sufficiently accurate models of all three receiver systems are
not yet available. In the case of the central beam of the Multibeam
receiver, existing models indicate that imperfections in the
orthogonality of the two feed probes may be as large as several
degrees \citep{vs04}. These imperfections break the fundamental
assumption of orthogonality made when applying a simple complex gain
correction and induce errors in the calculated total intensity of
order 1-2 \% \citep{osh+04}. Errors of this magnitude are only present
in the profile at phases where the fractional polarisation is close to
unity.

\section{Pulse Profile Stability} 
\label{sec:stability}

\begin{figure} 
\includegraphics[scale=0.35,angle=270]{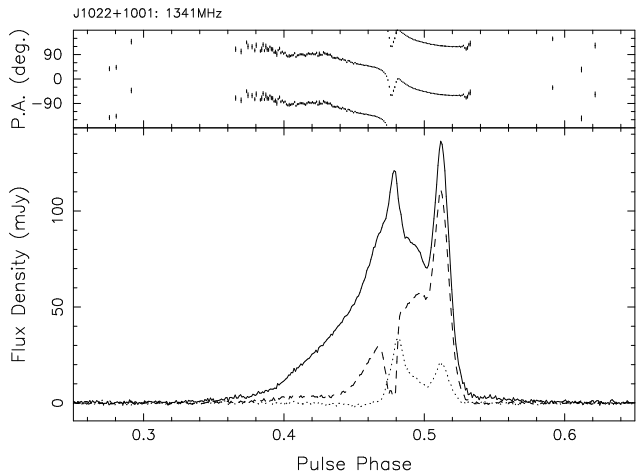}
\caption{PSR J1022+1001 average profile at a wavelength of 20 cm, formed
from data taken with the central beam of the Parkes Multibeam receiver.
The profile in the lower panel has been calibrated for relative 
instrumental gain and phase. It has a centre frequency of 1341.0 MHz 
and covers a bandwidth of 64 MHz. The solid curve represents total 
intensity (Stokes I), the dashed curve represents linearly polarised 
emission and the dotted curve represents circularly polarised emission. 
Note the high degree of linear polarisation in the trailing component. 
The relative (not absolute) position angle of the linearly polarised 
radiation is shown in the top panel. This profile has been calibrated 
against the flux of Hydra A (3C218).}
\label{fig:20std}
\end{figure}

\begin{figure} 
\includegraphics[scale=0.35,angle=270]{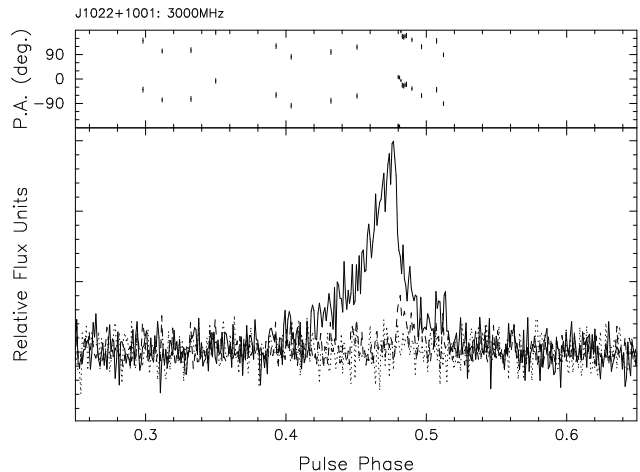}
\caption{PSR J1022+1001 average profile at a wavelength of 10 cm, 
formed from data taken with the new coaxial 10/50 cm receiver system
at Parkes. Again, the  profile has been calibrated for relative gain 
and phase. It has a centre frequency of 3000.0 MHz and covers a bandwidth 
of 64 MHz. Note the small degree of polarisation and the complete absence 
of the trailing pulse component. Flux calibrator observations were 
not made at the time these data were recorded, thus the amplitude scale 
is arbitrary.}
\label{fig:10std}
\end{figure}

\begin{figure} 
\includegraphics[scale=0.35,angle=270]{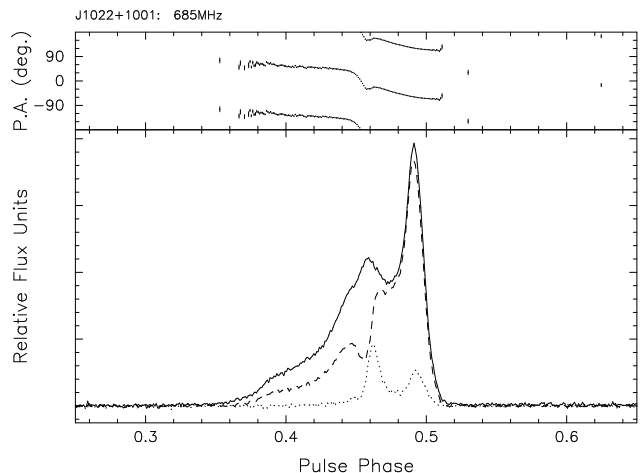}
\caption{PSR J1022+1001 average profile at a wavelength of 50 cm, 
formed from data taken with the new coaxial 10/50 cm receiver system 
at Parkes. Again, the profile has been calibrated for relative gain 
and phase. It has a centre frequency of 685.0 MHz and covers a bandwidth 
of 64 MHz. Note that the trailing component is almost 100 \% linearly 
polarised and significantly stronger than the leading component. Flux 
calibrator observations were not made at the time these data were recorded, 
thus the amplitude scale is arbitrary.}
\label{fig:50std}
\end{figure}

PSR J1022+1001 has an interesting profile morphology. At 20 cm the
profile consists of two sharp peaks (the principle components),
separated by about 0.05 phase units but joined by a bridge of emission
(Fig \ref{fig:20std}). This characteristic double-peaked shape was
critical to the analysis performed by \cite{kxc+99}, who make an
effort to characterise variations in the relative amplitude of the two
components by normalising against one or the other and computing
ratios of their height. The trailing pulse component is observed to
possess almost 100 \% linear polarisation \citep{osh+04, stc99,
xkj+98}, whilst the leading component corresponds to a local minimum
in the linear polarisation fraction which is near zero (Fig
\ref{fig:20std}).  The profile also changes rapidly as a function of
frequency, only the leading component is visible at a wavelength of 10
cm (Fig \ref{fig:10std}) and the amplitude ratio at 50 cm (Fig
\ref{fig:50std}) is nearly half that measured at 20 cm (Fig
\ref{fig:20std}). This spectral index variation within the profile and
the highly asymmetric nature of the polarised flux in the principle
components, combined with the fact that PSR J1022+1001 is prone to
strong episodes of scintillation, means that arrival time analysis
over broad frequency bands is easily subject to the introduction of
systematic errors. It also complicates any search for variability by
exacerbating the effect of instrumental imperfections, especially
those associated with polarimetry.

\subsection{Profile Normalisation}

In order to directly compare multiple integrated profiles from the
same pulsar, brightness changes due to interstellar scintillation must
be taken into account. This involves scaling or normalising each
profile by finding a characteristic value associated with the source
flux during the observation and adjusting the measured amplitudes to
ensure this value remains fixed across all observations. It should be
noted that applying a simple relative scaling to all observed profiles
masks any intrinsic variability in the total flux of the pulsar; but
that is not the subject of this paper. Using the PSRCHIVE
\citep{hvm04} development environment, software was constructed to
perform tests similar to those described by \cite{kxc+99}. In the
process of analysis and testing, we developed a normalisation scheme
based on the concept of difference profiles \citep{hmt75}.

\cite{kxc+99} describe a simple method of normalisation in which the
characteristic quantity associated with each integration is taken to
be the amplitude of one of the two component peaks. All the amplitudes
in the profile are scaled by the constant factor required to give the
chosen phase location the value of unity. In the case of PSR
J1022+1001, there is a choice as to which peak will be used as the
reference. An alternative characteristic quantity is the total flux
under the profile, or within a particular phase range. This quantity
is simply the sum of all the profile amplitudes within the region of
interest. Normalising by flux has the advantage that it uses no
morphological information (other than the duty cycle of the pulsed
emission in our case) and may therefore be better suited to detecting
subtle profile variations.

In this paper we compute fluxes after subtracting a baseline level
from each profile. A running mean with a phase width of 0.7 units is
used to estimate the baseline flux, which is then subtracted from each
bin. In order to further reduce our sensitivity to radio frequency
interference and other factors that can distort the uniformity of an
observed baseline, we normalise only by the flux in the on-pulse
region of the profile. This region is defined by an edge detection
algorithm that measures when the total flux crosses a threshold
level. This edge detection is performed on the standard template
profiles at each wavelength and the phase windows so defined are held
fixed for the remainder of the analysis. Unfortunately, normalising
profiles by flux tends to artificially amplify noisy observations
because a noise-dominated profile has a mean (and therefore a total
flux after baseline removal) approaching zero. In the data reduction
stage it is therefore advantageous to perform a cut on the basis of
S/N. This also helps reduce contamination by corrupted profiles. A S/N
cut of 100 was deemed the best compromise between retaining a large
fraction of the observed profiles and rejecting noise. This cut was
applied to the set of 5 minute CPSR2 integrations before any analysis
was commenced.

\cite{kxc+99}'s technique of normalising to the trailing feature has a
number of limitations:
\begin{itemize}
\item 
It cannot place any limits on the variability of the feature that
is used to normalise the profiles.
\item
Polarimetric calibration errors are maximised if the chosen component 
is highly polarised.
\item
Finite signal to noise profiles will vary due to noise even if there
is no intrinsic variability in the pulsar emission, this is true of
all components in the pulse.
\end{itemize}

\noindent We prefer our (more robust) flux-based normalisation scheme
which normalises by the total flux in the on-pulse region. Flux
normalisation allows us to properly identify which (if any) component
is varying as it makes no prior assumptions. It also reduces the
effect of poor calibration by averaging over the entire profile. There
will still be a background of variability due to noise, but it will be
spread uniformly.

\subsection{The Difference Profiles}

In order to detect subtle changes in pulse shape, it is possible to
subtract the amplitudes of a high S/N standard template profile from
an appropriately scaled and aligned copy of a given integration. This
procedure should yield Gaussian noise if the observed profile
precisely matches the standard; any morphological deviations will
protrude above the resulting baseline. Before differences can be
computed however, the observed profiles must be normalised to a
standard template. We choose to normalise all profiles according to
the amount of flux in the on-pulse region, as described above. In
addition, cross-correlation methods were used to determine relative
shifts, after which the observed profiles were rotated to align with
the standard template. Because profile variability studies are
sensitive to errors in polarimetric calibration, the difference
profile analysis was performed on both the calibrated and uncalibrated
profiles.  Only the 20 cm and 50 cm wavelengths had a sufficient
number of high S/N profiles to make a difference profile study
possible.

\begin{figure} 
\includegraphics[scale=0.35,angle=270]{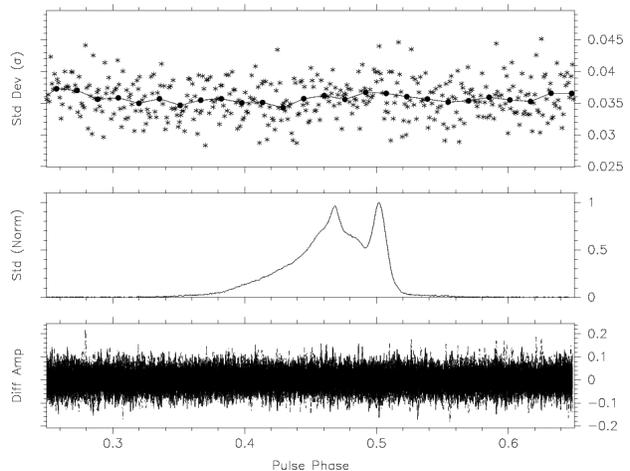}
\caption{The bottom panel shows superposed morphological difference 
profiles of 97 five minute total intensity (uncalibrated) integrations 
at a sky frequency of 1405.0 MHz. The middle panel shows the mean profile. 
The top panel shows the standard deviation of each phase bin (stars) in 
the difference profiles displayed in the bottom panel and a series 
(filled circles) representing the mean value of the standard deviation in 
64 windows, connected by lines, to aid perception of trends in the data.}
\label{fig:m_noc}
\end{figure}

\begin{figure} 
\includegraphics[scale=0.35,angle=270]{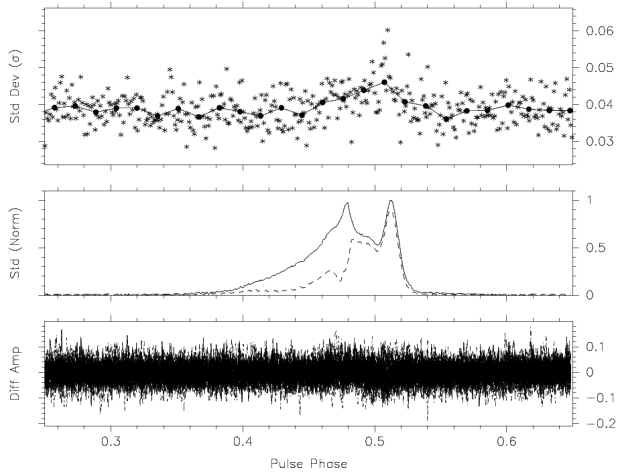}
\caption{As in Fig \ref{fig:m_noc}, superposed morphological difference 
profiles of 50 total intensity (calibrated) integrations at a sky frequency 
of 1405.0 MHz. The dashed line under the solid curve in the middle
panel represents the total polarised emission in the mean profile.}
\label{fig:m_cal}
\end{figure}

\begin{figure} 
\includegraphics[scale=0.35,angle=270]{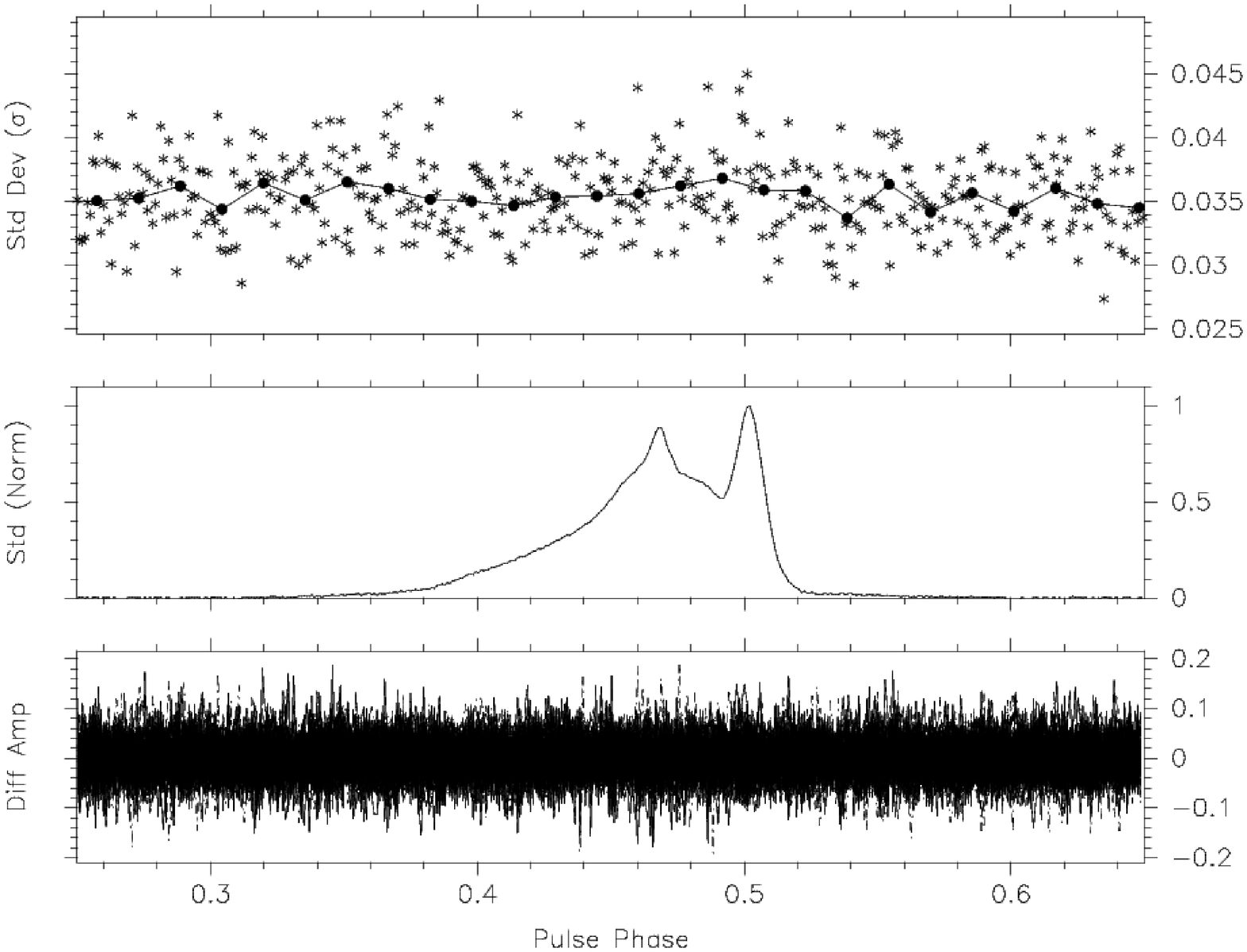}
\caption{As in Fig \ref{fig:m_noc}, superposed morphological difference 
profiles of 102 total intensity (uncalibrated) integrations at a sky 
frequency of 1341.0 MHz}
\label{fig:n_noc}
\end{figure}

\begin{figure} 
\includegraphics[scale=0.35,angle=270]{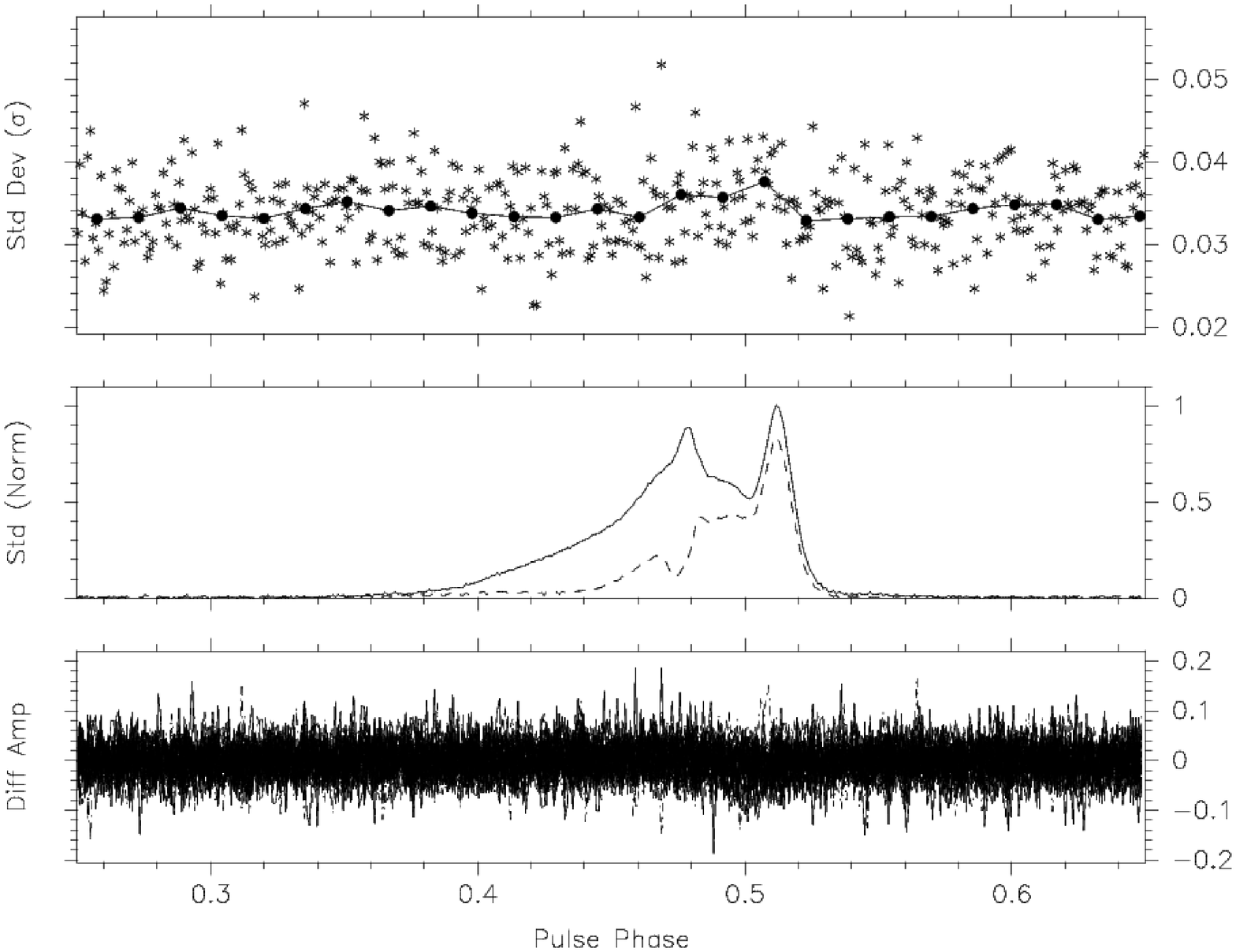}
\caption{As in Fig \ref{fig:m_cal}, superposed morphological difference 
profiles of 37 total intensity (calibrated) integrations at a sky frequency 
of 1341.0 MHz}
\label{fig:n_cal}
\end{figure}

\begin{figure} 
\includegraphics[scale=0.35,angle=270]{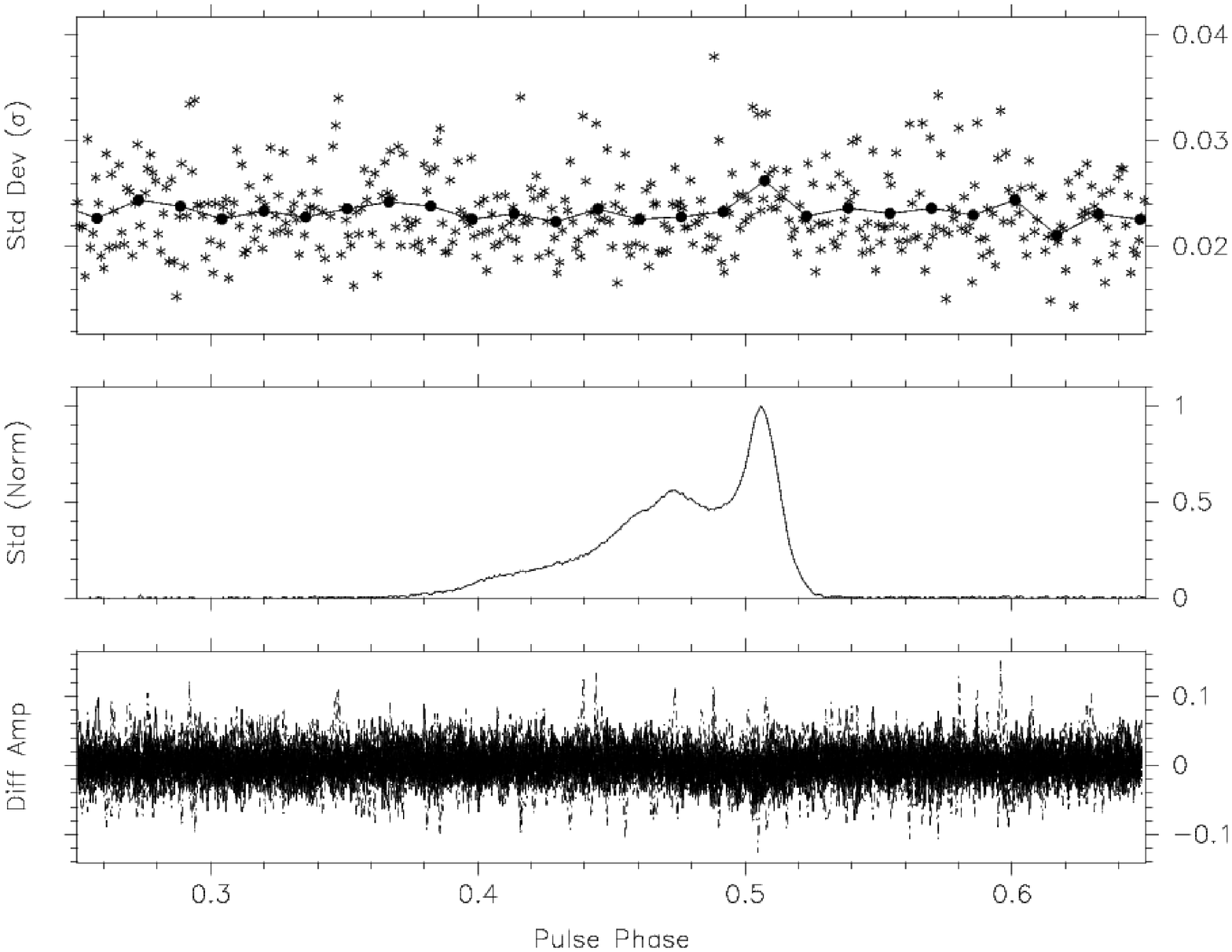}
\caption{As in Fig \ref{fig:m_noc}, superposed morphological difference 
profiles of 31 total intensity (uncalibrated) integrations at a sky 
frequency of 685.0 MHz}
\label{fig:50_noc}
\end{figure}

\begin{figure} 
\includegraphics[scale=0.35,angle=270]{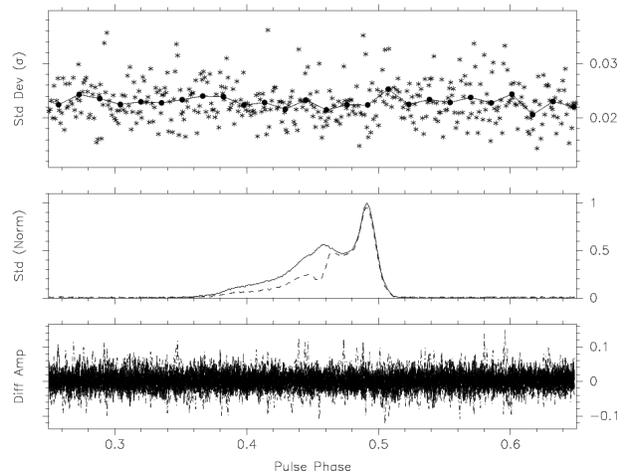}
\caption{As in Fig \ref{fig:m_cal}, superposed morphological difference 
profiles of 31 total intensity (calibrated) integrations at a sky frequency 
of 685.0 MHz}
\label{fig:50_cal}
\end{figure}

Figs \ref{fig:m_noc} - \ref{fig:n_cal} show difference profiles
computed from the CPSR2 observations at a wavelength of 20 cm, while
Figs \ref{fig:50_noc} \& \ref{fig:50_cal} show the 50 cm
results. The number of calibrated profiles is less than or equal to
the corresponding number of uncalibrated profiles, as not all
observations have corresponding calibration tracks.

As all figures show, the difference profiles are almost consistent
with noise, with no alarming peaks near the leading or trailing
components. This can be seen from both the difference profiles and
their standard deviations. The only evidence for variability comes
from Fig \ref{fig:m_cal} where $\sim$2 \% variability above the noise
floor can be seen near the trailing component. However, these
variations are not present in Fig \ref{fig:m_noc}, which represents
the same data set in the absence of polarimetric calibration. It would
seem that the calibration procedure itself is capable of inducing
profile instability, which is perhaps not surprising considering the
simplicity of the model used to correct for receiver imperfections.
In addition, CPSR2 has dynamic level setting that ensures the mean
counts are equal in both polarisations.  Provided the effective system
temperature in the two polarisations is similar, calibration is almost
unnecessary to form an accurate total intensity profile.  It therefore
seems a bizarre coincidence that these variations could be removed by
simply neglecting to calibrate the data, and that they correspond to
the maximum polarisation fraction in the profile. The simplest
interpretation is that across 64 MHz bandwidths, this pulsar's mean
profile is stable and that we are simply seeing the limitations of our
calibration procedure.

\subsection{Peak Ratio Evolution}

\cite{kxc+99} infer the presence of smooth variations in the relative
amplitudes of the principle components on time scales of a few minutes
to an hour or more by calculating peak amplitude ratios and
demonstrating that they evolve at a level significantly above the
uncertainty in the measurement. They also note that instances of such
smooth variation may not be common and that the time scales involved
can change from one data set to another.

\begin{figure}
\includegraphics[scale=0.3,angle=270]{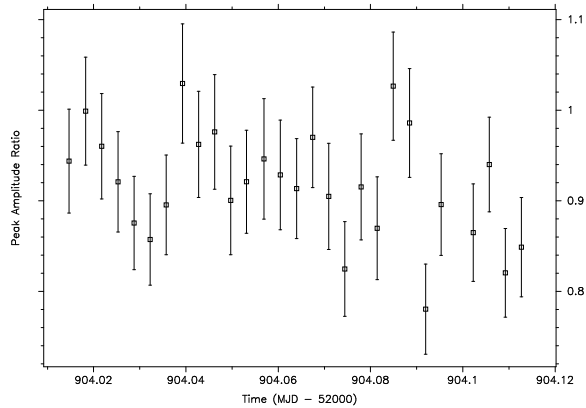}
\caption{Ratios of leading to trailing peak amplitudes during a
single 2.5 hour observation taken on September 22, 2003, at a frequency
of 1341.0 MHz. The error bars are computed based on the assumption that
the amplitude of each peak could vary by the RMS of the noise in the
baseline of the profile.}
\label{fig:ratio}
\end{figure}

As a secondary check, we computed similar amplitude ratios to that of
\cite{kxc+99} for our high S/N profiles. A S/N cut of 30 was deemed
sufficient for this analysis because it is less susceptible to
baseline corruption than the difference profile test. The RMS of the
off-pulse region in each profile was used as an indication of the
error in peak amplitude. The ratio of leading to trailing component
amplitude was computed for all profiles and plotted against
observation time (see Fig \ref{fig:ratio} for an example). In
addition, mean ratios and their associated standard deviation were
computed over the 15 month time span of our data set. These values are
presented in Tables 1 and 2.

\begin{table}
\begin{center}
\begin{tabular}{c|c|c}
\hline
Frequency (MHz) & Ratio  & Std Dev \\
\hline
1405.0 & 0.97 & 0.07 \\
1341.0 & 0.91 & 0.07 \\
685.0  & 0.58 & 0.02 \\
\hline
\label{tab:cal}
\end{tabular}
\caption{Ratio of leading to trailing component for the calibrated
data set.}
\end{center}
\end{table}

\begin{table}
\begin{center}
\begin{tabular}{c|c|c}
\hline
Frequency (MHz) & Ratio & Std Dev \\
\hline
1405.0 & 0.97 & 0.06 \\
1341.0 & 0.90 & 0.06 \\
685.0  & 0.58 & 0.02 \\
\hline
\label{tab:uncal}
\end{tabular}
\caption{Ratio of leading to trailing component for the uncalibrated
data set.}
\end{center}
\end{table}

The evolution of the profile's components with frequency is quantified
in Tables 1 and 2. It is interesting to note that calibration of the
data increases the scatter in the measured ratios at higher
frequencies, this further supports the notion that the variability
observed in Fig \ref{fig:m_cal} and by \cite{kxc+99} is likely due to
errors in the simple model used for polarimetric calibration. This
does not seem to be the case at lower radio frequencies, as the
calibration procedure has no detectable influence on the computed
component ratios at 685.0 MHz. The 50 cm system may simply be better
suited to the receiver model used for calibration, however the
intrinsic scatter at this frequency is also much smaller so the effect
of calibration may be imperceptible. To investigate the evolution of
these component ratios in time, the data set was examined by eye in
the hope of finding clear indications of non-random evolution between
5 minute integrations. Fig \ref{fig:ratio} shows the observation
that exhibited the most variation. Calibration has little effect on
these points. The variation about the mean is small and perhaps
insignificant.

\subsection{Summary}

The CPSR2 data set suggests that across a bandwidth of 64 MHz, PSR
J1022+1001 does not vary significantly. Any instabilities, if present,
must also be transitory in nature and therefore extremely difficult to
characterise. It is possible that variability with a random or
quasi-periodic structure, or small ($<$ 64 MHz) characteristic
bandwidth might still be present, but if so it would appear to have
little effect on the profile when all frequency channels are
combined. Profile instabilities should also be reflected in the timing
of the pulsar, which we investigate in the next section.

\section{Arrival Time Analysis}
\label{sec:timing}

\begin{figure}
\includegraphics[scale=0.35,angle=270]{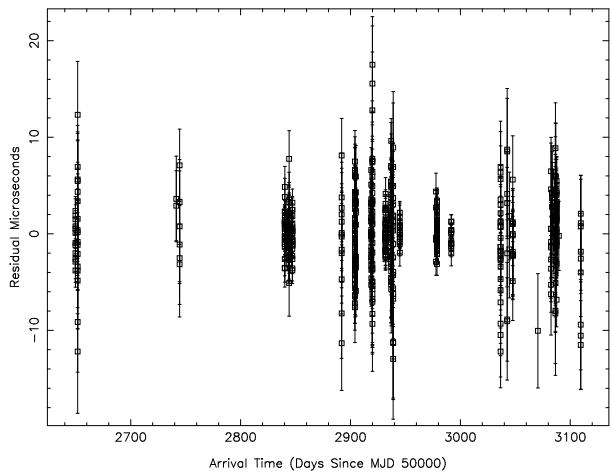}
\caption{Timing residuals from all 5 minute integrations with 
S/N in excess of 30 and timing error less than 7 $\mu$s. The 
RMS residual is 2.27 $\mu$s and the reduced $\chi^{2}$ of the 
weighted fit is 1.43. The vertical scale is an order of magnitude
better than arrival times presented by \cite{kxc+99} who saw
residuals as large as 400 $\mu$s.}
\label{fig:residuals}
\end{figure}

\cite{kxc+99} report that arrival times recorded at the Arecibo,
Jodrell Bank and Effelsberg radio telescopes yield a best fit RMS
residual of 15-20 $\mu$s when applied to a model of PSR J1022+1001 and
the binary system in which it resides. To achieve this RMS residual, a
complicated standard template profile consisting of five Gaussian
components with floating amplitudes was used, to compensate for
supposed intrinsic profile variability. In contrast, the CPSR2 data
set yields arrival times (from each 5 minute integration, across
multiple radio frequencies) that fit our timing model with a RMS
residual of 2.27 $\mu$s. These times of arrival (TOAs) were obtained
using a Fourier domain cross correlation process. Simple, static
standard template profiles were constructed from the sum total of
multiple integrations and the baseline noise was flattened to zero, to
decrease spurious self-correlation of the timing profiles with the
noise in the standard, which can lead to an RMS that is artificially
low. Separate standards were used for each frequency band and were
aligned to a common fiducial point. Our 5 minute (uncalibrated)
integrations were fit to a model of PSR J1022+1001 using the standard
pulsar timing package
TEMPO\footnote{http://www.atnf.csiro.au/research/pulsar/tempo/}.
Fig \ref{fig:residuals} shows the corresponding timing
residuals. It is interesting to note that although the front-end
receiver system was changed mid-way through the data set, there does
not seem to be any large systematic offset between the two receivers
and no jumps were used to fit across the boundary. Changes in cable
length or amplifier response between the two systems must have some
impact on the assignment of arrival times, however it would appear
that the offset is too small to measure in this data set.

A binary model of the type described by \cite{bt76} was used to model
the spin-down characteristics of PSR J1022+1001 and the perturbations
introduced by its companion. A list of fitted parameters and their
corresponding values (with error estimates) is presented below (Table
3). The errors in each TOA were uniformly scaled by a factor of 1.195
before fitting, to ensure the reduced $\chi^{2}$ per degree of freedom
was equal to unity. This is necessary to produce better error
estimates for the fitted parameters. Sky position is shown in both
Ecliptic and Equatorial (J2000) coordinates. PSR J1022+1001 lies in
the ecliptic plane, making it difficult to determine the ecliptic
latitude ($\beta$) accurately through timing measurements. This
accounts for the relatively high error in our measurement of $\beta$
in Table 3. In addition, the CPSR2 data has a time baseline of only 15
months, whereas the data presented by \cite{kxc+99} extends over more
than 4 years. As a consequence, we do not fit for a proper motion in
our model, choosing instead to hold this parameter fixed at the value
of --17 $\pm$ 2 mas yr$^{-1}$ quoted by \cite{kxc+99}. With proper
motion thus constrained, we obtain a significant estimate for the
parallax of the system. Given the short temporal baseline of our data
set, this parallax estimate should be considered preliminary and it is
included because it appears significant.

\hspace{2cm}

\begin{table}
\begin{center}
\begin{tabular}{l|l}
\hline
Parameter & Value \\
\hline
Ecliptic Lon. ($\lambda$) (deg)            & 153.86589029 (4)     \\
Ecliptic Lat. ($\beta$) (deg)              & -0.06391 (6)         \\
Proper Motion in $\lambda$ (mas yr$^{-1}$) & -17 (2) $^{*}$       \\
Parallax (mas)                             & 3.3 (8)              \\
\hline
Period (ms)                                & 16.4529296931296 (5) \\
Period Derivative ($10^{-20}$)             & 4.33 (1)             \\
Period Epoch (MJD)                         & 52900                \\
\hline
Dispersion Measure (cm$^{-3}$pc)           & 10.25180 (7)         \\
\hline
Projected Semi-Major Axis (lt-s)           & 16.7654148 (2)       \\
Eccentricity                               & 0.00009725 (3)       \\
Time of Periastron Passage (MJD)           & 52900.4619 (3)       \\
Orbital Period (days)                      & 7.805130160 (2)      \\
Angle of Periastron (deg)                  & 97.73 (1)            \\
\hline
Right Ascension ($\alpha$)                 & 10:22:58.015 (5)     \\
Declination ($\delta$)                     & +10:01:53.2 (2)      \\
\hline
Number of TOAS                             & 555                  \\
Total $\chi^{2}$                           & 545.74               \\
RMS Timing Residual ($\mu$s)               & 2.27                 \\
\hline
MJD of first TOA                           & 52649                \\
MJD of last TOA                            & 53109                \\
Total Time Span (days)                     & 460                  \\
\hline
\label{tab:eph}
\end{tabular}
\caption{PSR J1022+1001 \cite{bt76} timing model parameters 
derived from 15 months of CPSR2 observations. The error in the 
last significant digit is given in parentheses after the value.
($^{*}$ This value was given by \cite{kxc+99})}
\end{center}
\end{table}

\hspace{2cm}

Fig \ref{fig:residuals} shows that PSR J1022+1001 has benefited
from more regular observations in recent months. There are only a few
small groups of points in the earliest part of the data set. This is
another reason for not including proper motion in the model, as the
small number of points observed during the beginning of 2003 would
unreasonably dominate the fit. Parallax measurements are sensitive to
day of year coverage more than total observing time, so it is still
possible that the value of 3.3 $\pm$ 0.8 mas is meaningful. It is
interesting to note that our parallax measurement implies a distance
of only 300 $^{+100}_{-60}$ pc, as opposed to the value of 600 pc
derived from the \cite{tc93} electron density model. Many pulsars
within 1 kpc of the Sun are inconsistent with the \cite{tc93} model at
this level. If the system is indeed only 300 pc from the sun, other
secular changes may soon be detectable in the orbital parameters, such
as $a\sin i$.

\begin{figure}
\includegraphics[scale=0.35,angle=270]{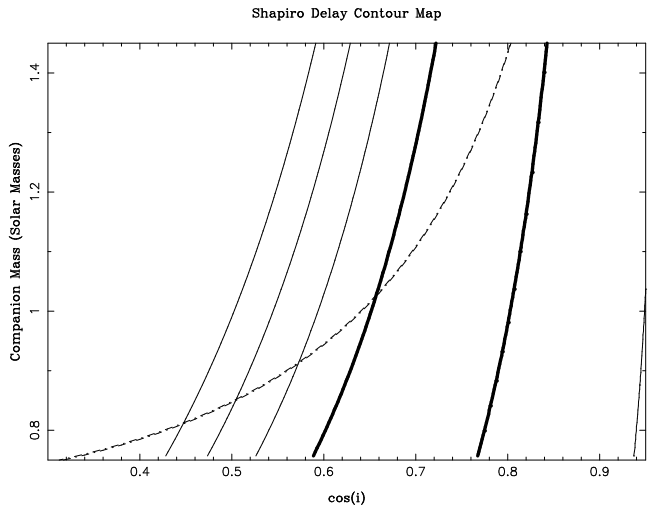}
\caption{Shapiro delay $\chi^{2}$ map. The thick lines represent
1-$\sigma$ contours, followed by 2,3 and 4-$\sigma$ contours on
the left hand side. The dashed line represents the mass function
constraint, assuming a 1.35 M$_\odot$ neutron star.}
\label{fig:shapiro}
\end{figure}

The accuracy with which we have been able to time this pulsar affords
the chance to derive new limits on several physical parameters. Of
particular interest is the possible presence of orbitally modulated
Shapiro delay as the distance each pulse must travel into the
companion star's gravity well changes throughout the 7.8 day
orbit. Whilst fitting for the Shapiro delay parameters $m_{\rm WD}$
and $i$ is not directly possible due to the small amplitude of their
timing signature, a statistical analysis of the effect these
parameters have on the $\chi^{2}$ of the fit can still reveal
important information. The parameters derived from our \cite{bt76}
model were incorporated into a model \citep{dd85,dd86} that includes
the two Shapiro delay parameters. This new model was used to create a
$\chi^{2}$ map for varying companion masses and inclination angles
(Fig \ref{fig:shapiro}). This map allows us to place upper and lower
bounds on the likely inclination angle, although it does not tightly
constrain the companion mass. Based on 2-$\sigma$ Shapiro delay
contours, $\cos i >$ 0.56. Assuming a neutron star mass of 1.35
M$_\odot$ and a white dwarf companion (a valid assumption given the
small eccentricity of the system), the Chandrasekhar mass limit for
the companion constrains $\cos i <$ 0.8. These two limits combined
imply that the inclination angle lies between 37$^{o}$ and 56$^{o}$
and suggest that $m_{\rm WD} >$ 0.9 M$_\odot$. This suggests that the
initial binary was near the limit required to produce two neutron
stars, and that the white dwarf is composed of heavier elements (maybe
even ONeMg) than many of the millisecond pulsar companions which are
most often He white dwarfs.

The model parameters presented in \cite{kxc+99} include a value for
the projected semi-major axis of the orbit, $x = a\sin i = $
16.765409(2) s. Given that the reference epoch of our data set is some
7 years ahead of the corresponding \cite{kxc+99} epoch, it is possible
to compare our value of $x$ in the hope of detecting a significant
change as the proper motion of the system alters our line of sight to
the plane of the orbit. We find $\dot x$ = 0.8 $\pm$ 0.3 $\mu$s
yr$^{-1}$. \cite{sbm+97} show that:

\begin{equation}
\label{equ:tani}
\dot x = x \cot i (-\mu_{\alpha} \sin \Omega + 
\mu_{\delta} \cos \Omega) {\rm s s}^{-1}
\end{equation}

\noindent where $\mu_{\alpha}$ and $\mu_{\delta}$ represent the
components of proper motion in RA and Dec and $\Omega$ is the
longitude of the ascending node. Only the component of $\mu$ in
ecliptic longitude was measured by \cite{kxc+99}, but given our
constraints on the inclination angle $i$, the only two unknowns in the
equation are now $\mu_{\delta}$ and $\Omega$ . If VLBI measurements
could provide a value for $\mu_{\delta}$ (and perhaps confirm our
detection of parallax), we would be able to place limits on the angle
$\Omega$, further constraining the 3-dimensional orientation of the
orbit on the sky.

\begin{figure}
\includegraphics[scale=0.35,angle=270]{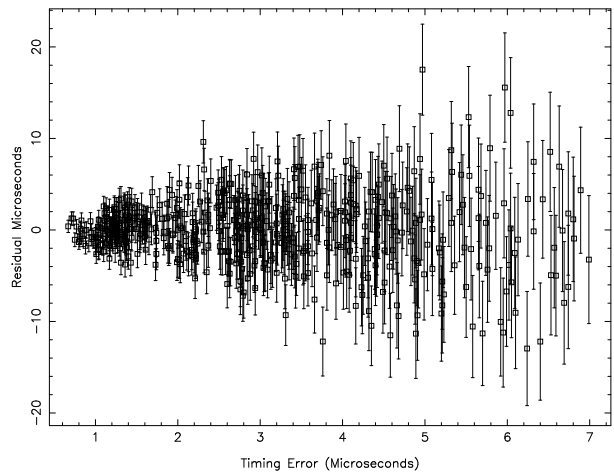}
\caption{Model residual plotted as a function of error in arrival
time.}
\label{fig:errors}
\end{figure}

The reduced $\chi^{2}$ of 1.43 indicates that the arrival times are
relatively free from unmodeled systematic effects. Increasing the
error estimates by 20 \% gives a $\chi^{2}$ of unity, which is low by
precision pulsar timing standards. Another way to view this low level
of systematics is to plot the residual against the arrival time
measurement uncertainty for each TOA (Fig \ref{fig:errors}). With
perhaps one exception, there are almost no points many standard
deviations from zero.

\cite{kxc+99} argue that the leading component of PSR J1022+1001
varies across a characteristic bandwidth of approximately 8 MHz. In
order to test this, a separate timing analysis was performed on the 20
cm wavelength data before complete summation of the frequency
channels, reducing the bandwidth in each integrated profile to 8 MHz.
If the pulsar is varying on these bandwidths and 5 minute time scales,
this should lead to an enormous increase in the RMS residual that
grossly exceeds the expected factor of $\sqrt{8}$. We find that the
RMS residual of these 8 MHz bandwidth arrival times is 6.65 $\mu$s,
which is very close to $\sqrt{8} \times 2.27 = 6.42$ $\mu$s, implying
that the increased scatter is consistent with the loss of S/N
associated with the reduction in bandwidth. This analysis is strong 
evidence that there are no profile variations with a characteristic 
bandwidth of approximately 8 MHz and time scales of a few minutes.

\section{Conclusions}
\label{sec:conclusions}

PSR J1022+1001 has an interesting pulse profile morphology in terms of
characteristic shape, polarimetric structure and evolution with radio
frequency. These factors conspire to make it a difficult source to
calibrate and analyse. Extensive studies of the morphological
differences between 5-minute integrations observed with CPSR2 at the
Parkes 64 m radio telescope are consistent with a stable pulse
profile. This bodes well for the future of precision timing of both
this source and millisecond pulsars in general. Although the RMS
residual presented is probably not small enough to make an immediate
contribution to any pulsar timing array, lengthier integrations and
continued monitoring may yet push the timing of this pulsar below the
1 $\mu$s mark.

We have used our improved timing to place some interesting limits on
the geometry of this source which demonstrate that the inclination
angle lies within the range $37^o < i < 56^o$. If our parallax is
confirmed, this system will make a good target for VLBI observations
which could both improve the distance to the pulsar and determine the
as yet unknown component of proper motion in the direction of ecliptic
latitude. This would provide additional limits on the three
dimensional orientation of the orbit via consideration of the change
in projected semi-major axis. In the future, limits on $\dot x$ will
improve and on a 10-year baseline, other relativistic observables may
become measurable.  We anticipate that with another 20 months of
timing, we should be able to obtain a more meaningful parallax and
independent proper motion for this source in right ascension. Our
error in $\omega$ is only 0.01$^o$, which suggests that the expected
rate of advance of periastron for this source $\dot \omega \sim 0.01^o
{\rm yr}^{-1}$ may be measurable on a 10-year baseline.

\vspace{0.5cm}

{\bf \noindent Acknowledgments.}

Parkes radio telescope is operated by the Australia Telescope National
Facility on behalf of the CSIRO. We thank Haydon Knight for observing
assistance. AWH is the recipient of an APA and CSIRO postgraduate
top-up allowance and thanks Claire Trenham for support and
encouragement.

\bibliographystyle{mn2e}
\bibliography{journals,psrrefs,modrefs}

\end{document}